\documentclass[12pt]{article}
\usepackage{epsfig,latexsym}
\def\f{\phi}
\def\t{\tau}
\def\th{\theta}
\def\ra{\rightarrow}
\def\o{\omega}
\def\e{\varepsilon}

\newcommand{\be}{\begin{equation}}
\newcommand{\ee}{\end{equation}}
\newcommand{\bea}{\begin{eqnarray}}
\newcommand{\eea}{\end{eqnarray}}

\def\a{\alpha}
\def\b{\beta}

\def\none{${\cal N}=1$ }
\def\ntwo{${\cal N}=2$ }
\begin{document}
\begin{titlepage}

\hfill IC/2003/50
\par\hfill CPHT-RR-034-0703\\
\vskip .1in \hfill hep-th/0307055

\hfill

\vspace{20pt}

\begin{center}
{\Large \textbf{Strings on pp-waves and Hadrons}}
{\Large \textbf{in (softly broken) ${\cal N}=1$ gauge theories}}
\end{center}

\vspace{6pt}

\begin{center}
\textsl{R. Apreda ${}^{a,b}$, F. Bigazzi ${}^{c}$ and A. L. Cotrone ${}^{d,e}$} \vspace{20pt}

\textit{a) Dipartimento di Fisica, Universit\`{a} di Milano-Bicocca, Piazza della Scienza, 3; I-20126 Milano, Italy.}\\
\textit{b) Dipartimento di Fisica, Universit\`{a} di Pisa, via Buonarroti, 2; I-56127 Pisa, Italy.}\\
\textit{c) The Abdus Salam International Centre for Theoretical Physics, Strada
Costiera, 11; I-34014 Trieste, Italy.}\\
\textit{d) Centre de Physique Th\'eorique, \`Ecole Polytechnique, 48 Route de Saclay; F-91128 Palaiseau Cedex, France.}\\
\textit{e) INFN, Piazza dei Caprettari, 70; I-00186  Roma, Italy.}\\
\end{center}

\vspace{12pt}

\begin{center}
\textbf{Abstract }
\end{center}

\vspace{4pt} {\small \noindent
We study the Penrose limit of Type IIB duals of softly broken ${\cal N}=1$ $SU(N)$ gauge theories in four dimensions, obtained as deformations of the Maldacena-N\`u\~nez and Klebanov-Strassler backgrounds. We extract the string spectrum on the resulting pp-wave backgrounds and discuss some properties of the conjectured dual gauge theory hadrons, the so called ``Annulons''. The string zero-point energy on the light-cone is nontrivial, due to the loss of linearly realized worldsheet supersymmetry, and negative, even in the unbroken supersymmetric case. This causes the appearance of non-perturbative corrections to the hadronic mass spectrum. We briefly discuss the thermodynamic behavior of these string models, calculating the corresponding Hagedorn temperatures.}
\vfill
\vskip 5.mm
 \hrule width 5.cm
\vskip 2.mm
{\small
\noindent
apreda@df.unipi.it\\
bigazzif@ictp.trieste.it\\
Cotrone@cpht.polytechnique.fr}

\end{titlepage}

\setcounter{page}{1}
%%%%%%%%%%%%%%%%%%%%%%%%%%%%%%%%%%%%%%%%%%%%%%%%%%%%%%%%%%%%%%%%%%%%%%%%%%%%%%%%%%%%%%%%%%%%%%%%%%%%
\section{Introduction}
A lot of work has been done to extend and test the $AdS/CFT$ correspondence \cite{maldaetal}, i.e. the conjectured equivalence between Type II string theory on $AdS_{d+1}\times M_{9-d}$ backgrounds and conformal field theories in $d$ dimensions. The extensions try to work out the equivalence in the non-conformal (and non-supersymmetric) case (for reviews see \cite{noirev}). The main checks of the correspondence have been provided for the simplest case of $d=4$, ${\cal N}=4$ SYM and Type IIB string theory on $AdS_5\times S^5$ with the RR five-form field strength turned on. No one knows how to exactly solve IIB strings on this background, therefore the checks at our disposal have been done exploring some particular simplifying limit of the original model. A first class of tests is limited to the low energy regime of the string model, i.e. to its supergravity approximation. A second class, instead, explores the correspondence at the stringy level, by considering IIB strings on a Penrose limit \cite{penrose} of the $AdS_5\times S^5$ background. This limit gives a maximally supersymmetric pp-wave background \cite{blau} on which IIB string is exactly solvable \cite{metse}. In \cite{BMN} this fact was used to find a precise map between the string-on-pp-wave modes and a sector of gauge invariant operators of the dual field theory. A perfect agreement was found, at least for a class of operators, between the stringy predictions and the gauge theory results.

Some successful attempts to extend the latter results to other conformal gauge theories have been done also in the non supersymmetric case \cite{noi0,klebetal}, finding sensible agreement between the field theory and the stringy theoretical predictions. In the non-conformal case some results have been obtained too, but the comparison between quantities on the two sides of the correspondence is  hard to perform, though the working philosophy is the same. One starts by considering supergravity solutions dual to some non-conformal gauge theory, then implements a Penrose limit which reduces the original background to some pp-wave solution on which, hopefully, string theory can be solved. This general procedure was applied in \cite{pandostras} to the (IR limit of the) Klebanov-Strassler (KS) \cite{ks} and the Maldacena-N\`u\~nez (MN) \cite{mn} regular supergravity solutions, which are conjectured to be dual to ${\cal N}=1$ $SU(N)$ Yang-Mills theory in four dimensions coupled to massive fields in the adjoint representation. The IIB string spectrum on the corresponding pp-wave backgrounds was obtained. From the gauge theory point of view, the Penrose limit procedure means considering a gauge theory sector made of hadrons of fixed, large mass in Minkowski space and carrying a large symmetry charge. The light-cone gauge string Hamiltonian describes their 3d non-relativistic motion as well as their excitations. 

In this paper we reexamine and extend the latter results to deformations of the MN and KS backgrounds which are conjectured to be dual to the above $SU(N)$ gauge theories after supersymmetry breaking. A class of regular solutions of this kind was found in \cite{gubsermn,gubserks} and particular solutions corresponding to the inclusion of a supersymmetry breaking scalar operator \cite{aharony}, or to a gluino mass term \cite{epz,apreda} were identified. Here we will be concerned with the softly broken regular solutions. After having pointed out the general philosophy underlying the Penrose limit procedure, fixing some problems of the MN limit in \cite{pandostras}, we examine the string spectrum on the resulting pp-wave backgrounds. Just as in the unbroken case these preserve 16 supercharges and so, as noted in \cite{cve}, no supersymmetry is linearly realized on the worldsheet. Thus the light-cone zero-point energy $E_0$ is non trivial, and should be calculated with some care, as pointed out in \cite{noistab}. In the cases at hand we show that $E_0$ is negative, even in the unbroken case (and contrarily to what was claimed in \cite{pandostras}). This however does not correspond to any (classical) instability as in the supergravity limit its value tends to a constant \cite{noistab}. In the opposite regime (and, again, also in the unbroken case), instead, its value tends to larger and larger negative values. Being this regime related to the weak effective coupling regime of the dual gauge theory sector, some consideration has to be done. The large negative contributions to the string Hamiltonian could be read as non-perturbative gauge theory corrections to the hadronic mass spectrum. These exist also in the unbroken supersymmetric case. After having pointed out the finite temperature behavior of our string models, we sketch the field theory part of the correspondence, concentrating our attention to the broken MN case (bMN in the following) being the broken KS one (bKS) very similar to its unbroken partner examined in detail in \cite{pandostras}. We are able to identify the hadrons dual to the zero mode part of the string spectrum in the ``universal sector'', the one determined by the symmetries of the supergravity solution.

The paper is organized as follows. In Section 2 we review the general way in which the Penrose limit of supergravity duals of gauge theories exhibiting confinement and mass gap is performed. In Section 3 we discuss the limit on the bMN background; we find the string spectrum on the corresponding pp-wave solution and discuss the properties of the zero-point energy. We also briefly examine the Hagedorn behavior of the corresponding finite temperature model, and find the expression for the critical temperature. The same is done for the more complicated bKS case in Section 4. In Section 5 we explore the field theory side of the correspondence, trying to extract predictions for the hadronic spectrum mainly in the bMN case.  
%%%%%%%%%%%%%%%%%%%%%%%%%%%%%%%%%%%%%%%%%%%%%%%%%%%%%%%%%%%%%%%%%%%%%%%%%%%%%%
\section{How to perform the Penrose limit}
%%%%%%%%%%%%%%%%%%%%%%%%%%%%%%%%%%%%%%%%%%%%%%%%%%%%%%%%%%%%%%%%%%%%%%%%%%%%%%
Let us consider the general form of the far IR (from the dual gauge theory point of view) limit of the (b)MN and (b)KS backgrounds. In both cases the metrics can be formally rewritten as
\be
ds^2= 2\pi\alpha'\,T_s dx_{\mu}dx^{\mu} +  {2\pi\alpha'\,T_s\over M_{KK}^2}ds_6^2 ,
\ee
where $T_s$ is the string tension and $M_{KK}$ is the mass of the Kaluza-Klein modes in the field theory.
The supergravity equations of motion only set ($M$ is the number of branes)
\be
{2\pi\alpha'\,T_s\over M_{KK}^2} = g_s\alpha'\,M.
\label{kk}
\ee
The value of
\be
T_s = {\sigma\over 2\pi\alpha'}
\ee
is instead not fixed. In the KS case, for example, $T_s = {\varepsilon^{2/3}\over g_sM\alpha'}$, but the definition of $\varepsilon$ is arbitrary: the only condition it has to satisfy is that $\varepsilon^{4/3}$ must have the dimensions of a length squared; since the only dimensionfull parameter in the low energy IIB theory is $\alpha'$, it is evident that $\varepsilon^{4/3}= c\,\alpha'$: however, nothing fixes the value of the number $c$. One could choose it as a pure number, or as a multiple of $g_sM$, and so on.

The Penrose limit we will take sends $T_s$ to infinity keeping $M_{KK}$ fixed \cite{pandostras}, and so the arbitrary $\sigma\rightarrow\infty$; at the same time the ratios of string tensions are fixed. The requirement that in the limit the KK and glueball masses remain fixed is also easily satisfied. From (\ref{kk}) we get
\be
M_{KK}^2 = {2\pi\alpha'\,T_s\over g_s\alpha'\,M} = {\sigma\over g_s\alpha'\,M},
\ee
which is fixed in the limit if we also take $g_s\,M\rightarrow\infty$. This is the usual Penrose limit appearing in all the cases considered in literature. No constraint or dangerous limit on $\alpha'$ or $g_s$ has to be taken.

%%%%%%%%%%%%%%%%%%%%%%%%%%%%%%%%%%%%%%%%%%%%%%%%%%%%%%%%%%%%%%%%%%%%%%%%%%%%%%
\section{The softly broken Maldacena-N\`u$\tilde{\rm{\bf{n}}}$ez model}
In this Section we study the Penrose limit of the softly broken MN solution \cite{gubsermn,epz} and the string theory on the resulting background.
The broken solution is in the same general form as the original MN one
\bea
\label{MNsol}
ds^2_{str} &= & e^{\Phi } \left[ dx_\mu dx^\mu
+ \alpha' N[ d \rho^2 + e^{ 2 g(\rho)}
(d\theta_1^2+ \sin^2\theta_1 d\phi_1^2)+
{1 \over 4 } \sum_a (w^a - A^a)^2 ] \right],
\\
G_3= ie^{\Phi}F_3& =& ie^{\Phi}\alpha'N \left[  -{1\over 4} (w^1 -A^1)\wedge (w^2 - A^2) \wedge ( w^3-A^3)  + { 1 \over 4}
\sum_a F^a \wedge (w^a -A^a) \right]
\eea
with a $\rho$-dependent dilaton whose value at the origin is a continuum parameter $\Phi_0$ and gauge field $A$ given by
\be
\label{Afield}
A={1\over 2} \left[ \sigma^1 a(\rho) d \th_1
+ \sigma^2 a(\rho) \sin\th_1 d\phi_1 +
\sigma^3 \cos\th_1 d \phi_1 \right].\ee
The one-forms $w^a$ are defined by
\bea
 { i \over 2} w^a \sigma^a &  = &dg g^{-1}, \qquad  \qquad\qquad  \qquad  \qquad \quad g = e^{ i \psi \sigma^3 \over 2 } e^{ i \th_2 \sigma^1 \over 2 }
e^{ i \phi_2 \sigma^3 \over 2}  ,          \nonumber \\
w^1 + i w^2 & = &e^{ - i \psi } ( d \th_2 + i \sin \th_2 d \phi_2)  ~,~~~~~~~~~~
w^3 = d \psi +\cos \th_2  d\phi_2.
\eea
The full explicit form of the functions $g(\rho),\, \Phi(\rho),\, a(\rho)\,$ is known only in the supersymmetric MN case. Otherwise, by simply looking at the supergravity equations of motion we can deduce their asymptotic behaviour.
We are going to study a Penrose limit of the softly-broken MN solution in the near-IR region, so we are concerned only with the $\rho\ra0$ behaviour of the functions above.
The asymptotics are \cite{gubsermn}
\begin{eqnarray}  
a(\rho) &=& 1 - b \rho^2 + ..., \nonumber\\    
e^g(\rho) &=&\rho - ( {b^2\over 4} +{1 \over 9}) \rho^3 + ...,\\   
\Phi(\rho)& =& \Phi_0 + ({b^2\over 4} + {1\over 3}) \rho^2 + ...,\nonumber    
\end{eqnarray}
where $b\in (0,2/3]$.
The range of allowed values for $b$ is imposed by requiring regularity of the supergravity solution and its linking with suitable UV asymptotics \cite{gubsermn}. The value $b=2/3$ corresponds to the supersymmetric MN solution.
The other values correspond to a non zero gaugino mass term in the dual filed theory, the function $a(\rho)$ being the supergravity dual of the gaugino bilinear \cite{noi1}.
The corresponding theory is then a non supersymmetric YM coupled to massive modes, among which there is the gaugino, in the adjoint of $SU(N)$.

Let us now shift the flat coordinates as $e^{\Phi_0/2}L^{-1}x_{\mu}\ra x_{\mu}$ where L is an arbitrary constant. In this way the tension for the confining strings of the dual gauge theory reads $T_s = L^2/(2\pi\alpha')$, while the glueball and KK masses\footnote{The decoupling of the 4-d YM theory from the KK modes is realized in the limit $e^{\Phi_0}N<<1$. This is beyond the validity of the supergravity approximation which instead requires $e^{\Phi_0}N>>1$ in order to have small curvatures.} are given by $M_{KK}^2\approx M_{gl}^2\approx L^2/(e^{\Phi_0}N\alpha')$.
Following \cite{pandostras}, we will take a Penrose limit of the IR of the supergravity background above, by enforcing the conditions
\be
L^2\approx e^{\Phi_0}N\ra\infty
\ee
which sends the string tension to infinity while keeping $M_{KK}, M_{gl}$ fixed.

To explore the IR of the above background it is convenient to perform a gauge transformation on $A$ such that it actually goes to zero when $\rho\ra0$. This can be done since $A$, also in the softly broken case, is a pure gauge in the extreme IR \cite{mn,epz}.
We use the gauge transformation \cite{bertmerl} $A\rightarrow h^{-1}A h+ih^{-1}dh$ with $h=e^{i\sigma^1\th_1/2}e^{i\sigma^3\phi_1/2}$.
The resulting field has the following expression in the $\rho\ra0$ approximation
\begin{eqnarray}
\label{AIR}
A &=& \Big(-{b\over 2}\rho^2 + {\mathcal{O}}(\rho^4)\Big)
\Big[\sigma^1(\cos\phi_1\,d\theta_1- \cos\th_1
\sin\th_1 \sin\phi_1\,d\phi_1 ) \nonumber\\
&&\; + \,\sigma^2(\sin\phi_1\,d\theta_1 + \cos\theta_1
\sin\theta_1 \cos\phi_1\,d\phi_1 ) +
\,\sigma^3(\sin^2\theta_1\,d\phi_1)\Big] .
\end{eqnarray}
Now we perform the Penrose limit on the ten dimensional background (\ref{MNsol})
along a null geodesic in the great circle on $S^3$ defined by $\theta_2=0$, $\phi_2=\psi$ and near $\rho\sim0$, and make the following change of variables
\begin{equation}
x^i \ra L x^i, \qquad \rho = {m_0\over L} \,r, \qquad \theta_2 = {2m_0\over L}\,v ,\qquad \phi_+ = {1\over
2}(\psi+\phi_2),
\end{equation}
where $e^{\Phi_0}\alpha'\, N= L^2/m_0^2$. This way we get a limit for the IR of the metric in (\ref{MNsol}), of the form
\begin{eqnarray}
ds^2 & = & -\,L^2dt^2 + \,dx_i dx^i
+\,dr^2 + r^2(d\th_1^2+\sin^2\th_1d \phi_1^2) \\
&&+ (dv^2+ v^2\,d\phi_2^2)
  + {L^2\over m_0^2}\,d\phi_+^2 - 2 v^2\,d\phi_2\,d\phi_+
  + b\,{r^2}\sin^2\theta_1\,d\phi_1d\phi_+ + {\mathcal{O}}(L^{-2}),\nonumber
\end{eqnarray}
where the new variables $r,v$ have dimension of length.

To reduce the metric in a more diagonal form, let us redefine \cite{pandostras}
\be
\hat \phi_1 = \phi_1+{b\over 2}\,\phi_+\qquad \hat \phi_2=
\phi_2-\phi_+.
\ee
We find
\begin{eqnarray}
ds^2 & = & L^2[-dt^2 + {1\over m_0^2}d\phi_+^2] + dx_i dx^i + {dr}^2 + r^2(d\theta_1^2+\sin^2\theta_1d\hat\phi_1^2) \nonumber\\
&&+ (dv^2+ v^2\,d\hat\phi_2^2)
   -( {v^2}+
   {{b^2r^2\over4}}\sin^2\theta_1)\,d\phi_+^2 + {\mathcal{O}}(L^{-2}).
\end{eqnarray}
Finally we define
\be
x^+=t, \qquad \qquad x^- = {L^2\over 2}(t- {1\over m_0}\,\phi_+),
\ee
and pass to the Cartesian coordinates $d u_1^2 + du_2^2 + dz^2= dr^2 +r^2( d\theta_1^2+\sin^2\theta_1d \hat \phi_1^2)$, $\,d v_1^2 + dv_2^2 = dv^2+ v^2\,d\hat\phi_2^2$
.  
So, we obtain
\be\label{ppMNr}
ds^2 = -2dx^+dx^- -m_0^2\,({b^2\over 4}u_1^2 + {b^2\over 4}u_2^2+v_1^2+v_2^2)(dx^+)^2 + 
d\vec{x}^{\,2} +
dz^{\,2} + du_1^2+du_2^2+dv_1^2+dv_2^2 \ .
\ee
The string action on this background will thus include four massless scalars ($x_i$ and $z$) and four massive ones, just as in the supersymmetric case. The only difference is in the value of the masses of the two scalars $u_1, u_2$ which are now $b$-dependent.
Note that the changing is restricted to the ``non-universal'' sector of the theory, i.e. the one which is non determined by the symmetries of the original background.
This is expected, since the soft-supersymmetry breaking term doesn't change the overall topology of the metric in the far IR.
As a consequence, the main features of the field theory annulons will be the same as in the supersymmetric theory.

The Penrose limit of the 3-form gives\footnote{It is easy to check that the background here obtained still satisfies the supergravity equations of motion, as ($g^{++}=0$) $R_{++}= {1\over4}(G_{+ij}G_{+}^{ij})$.}
\be\label{ppH}
 G_3 = -2 i m_0 dx^+\wedge [dv_1\wedge dv_2 + {b\over 2} du_1\wedge du_2].
\ee
The string Hamiltonian on the above background is
\be
\label{MNH}
 H = -p_+ = i\partial_+ = E - m_0 (-{b\over 2} J_1 + J_2 + J_{\psi}) \equiv
E-m_0\,J,
\ee
and the momentum $P^+$ is
\be
P^+ = - {1\over 2} p_- = {i\over 2}\partial_- = {m_0\over L^2}
(-{b \over 2} J_1 + J_2 + J_{\psi}) = {m_0\over L^2} J ,
\ee
where, following the notations in \cite{pandostras}, we denote $-i\partial_{\phi_1}$,
$-i\partial_{\phi_2}$ and  $-i\partial_{\psi}$ with $J_1$, $J_2$ and $J_\psi$ respectively.

%%%%%%%%%%%%%%%%%%%%%%%%%%%%%%%%%%%%%%%%%%%%%%%%%%%%%%%%%%%%%%%%%%
\subsection{String theory on the bMN pp-wave}
As the reader can see, the only difference between the results collected above and the ones obtained in the supersymmetric case \cite{pandostras} amounts in replacing the parameter b with $2/3$ in the $u_1, u_2$ sector. All the calculations done in \cite{pandostras} for the spectrum of the bosonic and fermionic worldsheet fields, as well as the expression for the string Hamiltonian, can be easily exported to our case.
Thus, studying the string action on the pp-wave background (\ref{ppMNr}), (\ref{ppH}), and choosing the light-cone gauge as usual ($x^+=\alpha'p^+\tau$), produces the following results.

Let us define $m=m_0 \alpha' p^+$. The bosonic sector of the system is described by four massless fields
($x^i, z$) with frequencies $\omega_n=n$, and four massive fields ($u_1,u_2,v_1,v_2$) with frequencies
\be
\omega_n^u = \sqrt{n^2 + {b^2\over 4}\,m^2},\qquad \qquad 
\omega_n^v = \sqrt{n^2+ m^2}.
\ee
The worldsheet fermions are instead all massive and their frequencies read
\bea
\omega_n^{I}&=& \sqrt{n^2 + {m^2\over 4}\Biggl(1+{b\over 2}\Biggr)^2}, \qquad I=1,2,3,4 \nonumber \\
\omega_n^{J}&=& \sqrt{n^2 + {m^2\over 4}\Biggl(1-{b\over 2}\Biggr)^2}, \qquad J=5,6,7,8 .
\eea
In the original metric (\ref{MNsol}) the coordinates $v_1,\, v_2\,$ represent the directions, on the 3-sphere, normal to the reference geodesic chosen for the Penrose limit, and,
together with the $x^i$, parameterize the so called ``universal sector'' \cite{pandostras}.
The soft breaking of the MN solution does not affect this sector, since it is determined by the deep IR symmetries
of the original background, i.e. by its deformed conifold shape.
The modes $u_1,\, u_2\,$ depend on the details of the solution.
They have different (zero-mode) masses, which depend on the supersymmetry breaking parameters.
All these features will be found also in the bKS case, where the mass of the $z\,$ mode too will depend
on the breaking parameters.
In the bMN solution, instead, it is massless in both the supersymmetric and the non supersymmetric case.

The sum of the squares of the fermionic frequencies above exactly
matches the sum of the squares of the frequencies of the bosonic
fields order by order in $n$.  Thus the corresponding string-theory is finite also in the $b\neq 2/3$ case. 

The construction of the string Hamiltonian proceeds as in \cite{pandostras} and we will not repeat it here.
The only point we would like to stress concerns the evaluation of the ground state energy, which turns out to be non zero.
In fact, this string theory does not have any linearly realized supersymmetry.
In the Penrose limit of Type IIB string theories one usually encounters some enhancement
of the original supersymmetry, even when the original theory didn't have any.
This of course does not happen in this case, because even the original unbroken theory had no
linearly realized supersymmetries.
It had in fact only the ubiquitous sixteen kinematical ones, which do not give
a supersymmetric spectrum \cite{cve}.

The zero point energy reads
\be \label{zeromn}
E_0(m)={m_0\over 2m}\sum_{n=-\infty}^{\infty}\left[ 4n + 2\omega_n^u + 2\omega_n^v - 4\omega_n^1 -4\omega_n^5 \right] .
\ee
We can calculate this finite sum using the Epstein function \cite{noistab}, defined as (see for example \cite{Albu})
\begin{eqnarray}\label{epstein}
F[z,s,m^2]&=& \sum_{n=1}^{\infty}\left[(n+s)^2 + m^2\right]^{-z}= {1\over2}[(s+1)^2 +m^2]^{-z} +
\int_1^{\infty}dx[(x+s)^2+m^2]^{-z} + \nonumber \\
&+& i\int_0^{\infty}dt\left[{ [(1+it+s)^2+m^2]^{-z}-
[(1-it+s)^2+m^2]^{-z}}\over{e^{2\pi t}-1}\right].
\end{eqnarray}
When $m\ra0$, $E_0$ only gets contributions from the zero frequencies, so
\be
E_0\ra {m_0\over2}[|b| +2 -2|1-{b\over2}| -2|1+{b\over 2}|] \leq 0 ,
\ee
the equality holding only for $|b|=2$, which is actually excluded by the condition $b\in (0,2/3]$. 
Thus, for all the allowed values of $b$, and actually also for the supersymmetric solution $b=2/3$, {\it the zero-point energy is negative for $m\ra0$}. It also stays negative for every value of $m$. In the large $m$ limit in fact, as suggested by the Epstein function and stressed in \cite{noistab}, the series over $n$ in (\ref{zeromn}) can be approximated by an integral over $x\in (-\infty,\infty)$ and we get
\be
E_0\ra -{m_0 m\over 4}\Biggl[2{b^2\over 4}\log\Biggl({b^2\over 4}\Biggr) - 4\Biggl({b\over 2}-1\Biggr)^2\log\Biggl({b\over 2}-1\Biggr)^2
-4\Biggl({b\over 2}+1\Biggr)^2\log\Biggl({b\over 2}+1\Biggr)^2\Biggr]. 
\label{Imn}
\ee
This depends linearly on $m$ and thus $E_0$ takes larger and larger negative values as $m$ increases.
This fact could be read as a signal of the existence of non perturbative corrections to the annulon mass spectrum.

While a negative light-cone energy can seem dangerous at first sight, in \cite{noistab} it was shown that whenever it goes, in the supergravity limit $m\ra 0$, to a constant value independent from $p^+$, the theory is classically stable, no matter the sign of $E_0$.
Moreover, it is not clear what kind of instability there could be in the dual field theory.
In the present case, the field theory by itself is surely stable in the supersymmetric case, and it was argued to be so also in the softly broken case \cite{aharony,epz}, if the value of the gaugino mass is small enough (i.e. if $b$ is not very different from $2/3$).
In the Penrose limit we are considering states in this theory which have large mass $E$ and large charge $J$.
We should then see an instability in $E-m_0J$ for these states and it is far from clear what it could be\footnote{We warn the reader that in the Type 0 case \cite{noi0} there is a classical instability \cite{noistab}, which should correspond to an instability in $\Delta -J$ in the dual field theory, $\Delta$ being the conformal dimension of the operators. Also in this case what the field theory instability would look like is unclear, being non perturbative in the effective coupling constant \cite{noi0}.}.
So we will proceed assuming that the theory is stable, at least for small $m$, being aware that the situation in the interacting quantum string theory could be different.

Finally, let us spend a few words on the thermodynamics of this string model.
Since the latter is exactly solvable, we can evaluate the string partition function at finite temperature
and search for the usual Hagedorn behavior.
The calculation is standard and straightforward\footnote{See, for example, \cite{greene}.}, and as usual the
Hagedorn temperature is given by the zero-point energy evaluated with anti-periodic fermions.
It is given implicitly by ($T_H=\beta_H^{-1}$)
\be
\beta_H= -2\sqrt{2}\pi\alpha' E_{0}\left({m_0\beta_H\over2\sqrt{2}\pi};0,1/2\right)
\label{hagmn}
\ee
where $E_{0}(M;0,1/2)$ is the bMN zero-point energy $E_0(M)$ (\ref{zeromn}) evaluated with periodic bosons and antiperiodic fermionic fields (i.e. with $s=1/2$ in formula (\ref{epstein})). Our result differs from the one in \cite{pandovaman} for a finite term in $E_{0}$: in fact the zero-point energy in \cite{pandovaman} is calculated in terms of a Casimir energy and this amounts on subtracting to the finite sum (\ref{zeromn}) the finite term (\ref{Imn}). However, as pointed out in \cite{noistab} this choice is not justified in superstring theory, where one has first to evaluate the physical quantities and then eventually ask for a regularization/renormalization prescription. If the physical quantities are finite, as in the present case happens to $E_0$, no ad hoc subtraction is allowed.

%%%%%%%%%%%%%%%%%%%%%%%%%%%%%%%%%%%%%%%%%%%%%%%%%%%%%%%
\section{The softly broken Klebanov-Strassler model}

In this case it is convenient to start with the (Einstein frame) supergravity fields written in the form \cite{bgz,apreda}
\begin{eqnarray}\label{ksback}
2^{-1/2}3^{-3/4}ds^2&=&\biggl[c\varepsilon^{4/3}e^{-5q(\t)+2A(\t)}dx_{\mu}dx^{\mu} + e^{3q(\t)-8f(\t)}ds_6^2\biggr],\nonumber \\
ds_6^2&=&{1\over9}(d\t^2+g_5^2) + {e^{10f(\t)+y(\t)}\over6}(g_1^2+g_2^2)+ {e^{10f(\t)-y(\t)}\over6}(g_3^2+g_4^2),\label{ksmet}\\
B_{(2)}&=&2g_sP \biggl[ g(\t)g_1\wedge g_2+k(\t) g_3\wedge g_4\biggr],\label{bidue}\\
F_{(3)}&=& 2P \biggl[g_5\wedge g_3\wedge g_4+d[F(\t)(g_1\wedge g_3+
g_2\wedge g_4)]\biggr],\\
F_{(5)}&=& 4g_sP^2 L(\t)g_1\wedge g_2\wedge g_3\wedge g_4\wedge g_5 +\nonumber\\
&& \quad + *\biggl[4g_sP^2 L(\t)g_1\wedge g_2\wedge g_3\wedge g_4\wedge g_5\biggr], \\
L(\t)&=& [k(\t)-g(\t)] F(\t) + g(\t), \label{elle}
\end{eqnarray}
where $\e$, the conifold deformation parameter, gives the mass scale of the theory $M_s\sim \e^{2/3}$, $c$ is a numerical constant and $P=M\alpha'/4$, where $M$ is the number of fractional branes on the conifold.  
We used the following basis of one-forms
\begin{equation}
g^1 = {{e^1 - e^3}\over\sqrt2}, \quad g^2 = {{e^2 - e^4}\over\sqrt2},\quad
g^3 = {{e^1 + e^3}\over\sqrt2}, \quad g^4 = {{e^2 + e^4}\over\sqrt2},\quad
g^5 = e^5,
\label{fbasis}
\end{equation}
with
\begin{eqnarray}
e^1= -\sin\theta_1d\phi_1, \qquad e^2= d\theta_1, \qquad e^3= \cos\psi\sin\theta_2d\phi_2 - \sin\psi d\theta_2, \nonumber \\
e^4= \sin\psi\sin\theta_2d\phi_2 + \cos\psi d\theta_2, \qquad e^5= d\psi + \cos\theta_1d\phi_1 + \cos\theta_2d\phi_2.
\end{eqnarray}
As in the bMN case, there is no analytic form for the various functions in (\ref{ksback})/(\ref{elle}) apart from the supersymmetric one. 
Nevertheless, we are interested only in the IR behavior of the fields, which reads \cite{apreda} 
\setlength\arraycolsep{2pt}
\begin{displaymath}
\begin{array}{cccclccccccccccc}
%\begin{displaymath}
\\[-2mm]
\texttt{\large{$A$}}(\t) &=&\; \frac{2}{3}\; \texttt{Log}(\t) &+& \frac{1}{6}\texttt{Log}(\frac{A_0}{32})&&&+&a_2 \t^2&&&&&&&+ \; \;\ldots \\[2mm]
\texttt{\large{$q$}}(\t) &=&\frac{4}{15} \; \texttt{Log}(\t) &+& \frac{1}{6}\texttt{Log}(A_0 \, 3^{\frac{9}{10}} \, 2^{-\frac{7}{5}})&&&+&q_2 \t^2&&&&&&&+ \; \;\ldots \\[2mm]
\texttt{\large{$f$}}(\t) &=&\frac{1}{10} \; \texttt{Log}(\t) &+& \frac{1}{10}\texttt{Log}(\frac{2}{3})&&&+&f_2 \t^2&&&&&&&+ \; \;\ldots \\[2mm]
\texttt{\large{$y$}}(\t) &=&\quad \; \texttt{Log}(\t) &-& \texttt{Log}(2)&&&+&Y\, \t^2&&&&&&&+ \; \;\ldots\\[2mm]
\texttt{\large{$\phi$}}(\t) &=&&\Phi_0&&&&+&\Phi_2 \t^2&&&&&&&+ \; \;\ldots\\[2mm]
\texttt{\large{$F$}}(\t) &=&&&&&&+&\texttt{\large{$F$}} \, \t^2&&&&&&&+ \; \;\ldots \\[2mm]
\texttt{\large{$k$}}(\t) &=&&&&+&K \t&&&+&k_3 \t^3&&&&& +\; \;\ldots \\[2mm]
\texttt{\large{$g$}}(\t) &=&&&&&&&&+&G \t^3&&&&& +\; \;\ldots\\[2mm]
\end{array}
\end{displaymath}
The string coupling is related with the value of the dilaton in zero $g_s=e^{\Phi_0}$.
This solution has five free\footnote{It is very difficult to find the range of these parameters allowing the linking of the IR with the UV asymptotic solutions, due to the lack of precision of the numerical integration of the equations of motion. We don't have a complete control of these data.} parameters: $\Phi_0,\,Y,\,F,\,G,\,K$. 
The supersymmetric solution corresponds to 
\be
Y=-1/12,\quad F=1/12,\quad K=1/3,\quad G=1/12,\quad \Phi_0=0. 
\label{sulim1}
\ee 
Moreover $A_0^{1/2}\approx e^{\Phi_0}M\a'$. 
The remaining interesting coefficients are
\bea
a_2 &=& {1\over10}+{4\over15}Y + {P^2e^{\Phi_0}\over A_0}[-{2\over9} +16G^2-{1\over3}K^2], \nonumber \\
q_2 &=& {1\over25}+{8\over75}Y +{P^2e^{\Phi_0}\over A_0}[-{11\over90}-{8\over5}F^2 + {8\over5}G^2-{7\over30}K^2], \nonumber \\
f_2 &=& -{7\over200}-{13\over50}Y+{P^2e^{\Phi_0}\over A_0}[{1\over10}-{8\over5}F^2 - {72\over5}G^2+{1\over10}K^2].
\label{intcoef}
\eea
This is a regular, stable, non supersymmetric deformation of the Klebanov-Strassler solution, including mass terms for the gauginos.

We are again interested in keeping the glueball masses fixed, i.e. fixed $M_{gb} \sim {\e^{2/3}\over e^{\Phi_0/2}A_0^{1/2}}$, while taking the Penrose limit. Thus we define
\be
L^2= {c\varepsilon^{4/3}\over A_0^{1/2}}, \qquad \qquad m_0^2= {L^2\over2A_0^{1/2}},
\ee
and we consider the limit $L\rightarrow\infty$, keeping $m_0$ fixed.
Again, the flux tube tensions are diverging in the limit $T_s \sim M_{gb}^2 (g_s M) \rightarrow \infty$.
The Penrose limit goes on exactly as in the bMN case, choosing a null geodesics at $\tau=0$ and spinning on the equator of the three-sphere in the background (\ref{ksback}).
We rescale $\t \ra \tau/L$, expand the metric in powers of $\tau/L$ and ignore terms which vanish in the $L \ra \infty$ limit. Let us first change angular variables $(\psi,\theta_1,\phi_1,\theta_2,\phi_2)\rightarrow (\psi',\theta,\phi,\theta',\phi')$  switching to a basis of one-forms $\o_1,\o_2,\o_3$ related to the $g_i$'s by \cite{pandostras}
\bea
g_5&=& \sin\theta\cos\phi\omega_1-\sin\theta\sin\phi\omega_2+\cos\theta\omega_3,\nonumber \\
\cos(\psi/2)g_1+\sin(\psi/2)g_2&=&{1\over\sqrt2}(\cos\theta\cos\phi\omega_1-\cos\theta\sin\phi\omega_2-\sin\theta\omega_3-2\sin\theta d\phi),\nonumber \\
-\sin(\psi/2)g_1+\cos(\psi/2)g_2&=&{-1\over\sqrt2}(\sin\phi\omega_1+\cos\phi\omega_2-2d\theta),\nonumber \\
\cos(\psi/2)g_3+\sin(\psi/2)g_4&=&{1\over\sqrt2}(\cos\theta\cos\phi\omega_1-\cos\theta\sin\phi\omega_2-\sin\theta\omega_3),\nonumber \\
-\sin(\psi/2)g_3+\cos(\psi/2)g_4&=&{-1\over\sqrt2}(\sin\phi\omega_1+\cos\phi\omega_2).
\eea
The one-forms $\o_i$ are given by $T^\dag\,dT = - dT^\dag\,T = {i\over 2}\,\omega_a\,\sigma^a,\,\, T = e^{{i\over 2}\,\phi'\, \sigma^3}\,e^{{i\over 2}\,\theta'\,
 \sigma^1}\,e^{{i\over 2}\,\psi'\, \sigma^3}$.

The limit is realized around $\theta'=0$, i.e. we rescale $\theta'\rightarrow\theta'/L$. Let us define $2\phi_+= \phi'+\psi'$. The Penrose limit on the metric (\ref{ksmet}) thus gives  
\bea
ds^2 &=& -L^2\left[1+(2a_2-5q_2)\frac{\t^2}{L^2}\right]\,dt^2+ L^2d\vec{x}^2 +  {1\over4m_0^2}\Biggl[ 4L^2d\f_+^2 + d\t^2 \nonumber \\
&& + \t^2(d\th^2 +\sin^2{\th}\,d\f^2)+ (d\th')^2 + (\th' d\f')^2  - 2(\th')^2d\f'd\f_+ +2\t^2\sin^2\theta d\phi d\phi_+ \nonumber \\
&&  + 4\t^2(3q_2-8f_2)\cos^2{\th}\, d\f_+^2 + 4\t^2(\frac{1}{4}+3q_2+2f_2-Y)\sin^2{\th}\, d\f_+^2 \Biggr].
\eea
The change of coordinates \cite{pandostras}
\be
 x^+ = t,
 \qquad x^{-}= {L^2\over 2} \left(t - {1\over m_0} \,{\phi_+ }\right),
\qquad \varphi = {1 \over 2 }(\phi' - \psi'), \qquad \tilde{\phi} = \phi
 + \,\phi_+,
\ee
\be
v = {1\over 2m_0}\,\theta'\,e^{i\varphi},
\qquad z = {1\over 2m_0} \,\tau\cos\theta ,
\qquad u = {1\over 2m_0}\,\tau\sin\theta\,e^{i\tilde{\phi}} , 
\ee
and the rescaling $x^i \ra x^i/L$,
give a pp-wave background
\be
\label{metks}
ds^2 = -4dx^+dx^- - m_0^2\big[\,\a^2 z^2 +
 \,\b^2 u\bar{u} + v\bar{v}\big] \, (dx^+)^2 
 + d\vec{x}^{\,2} + dz^2 + dud\bar{u} + dvd\bar{v},
\ee
with
\be
\a^2=\left(8a_2 - 32q_2 +32 f_2 \right),  \qquad\qquad  \b^2=\left(8a_2 - 32q_2 - 8 f_2 +4Y \right).
\ee
In terms of the free parameters we have
\bea
\a^2 &=& -{8\over5} -{48\over5}Y + {P^2e^{\Phi_0}\over A_0}[{16\over3} -384G^2 +8K^2], \nonumber \\
\b^2 &=&  -{1\over5} +{24\over5}Y + {P^2e^{\Phi_0}\over A_0}[{4\over3} +64F^2+ 192G^2 +4K^2].
\eea
In the supersymmetric case we get
\bea
\a^2 &=& -{4\over5} + {32\over9}{P^2e^{\Phi_0}\over A_0}, \nonumber \\
\b^2 &=& -{3\over5} + {32\over9}{P^2e^{\Phi_0}\over A_0},
\eea
which reproduce the results in \cite{pandostras} with 
\be
\gamma^2\equiv{P^2e^{\Phi_0}\over A_0}={9\over8}{a_1\over a_0}.
\label{sulim2}
\ee
The only difference with respect to the supersymmetric case is a change in the coefficients of the $z$ and $u,\, \bar{u}$ coordinates.
In the string theory it means that the masses for three world-sheet scalars depend on the free parameters of our theory.
As in the bMN case, the changing is restricted to the ``non-universal'' sector of the theory.

From formulae (\ref{bidue})/(\ref{elle}) it is straightforward to verify that the RR and NSNS forms get in the Penrose limit a very simple dependence on the free parameters, so that the complex-three form $G_3 = H_3 + ie^{\Phi} F_3$ reads\footnote{$\Phi \rightarrow \Phi_0$ in the limit.}
\be
\label{G3}
G_3 = -{2m_0e^{\Phi_0}P\over A_0^{1/2}}\,dx^+\wedge\left[
    (4F\,du\wedge d\bar{u}+  \, dv\wedge d\bar{v})
    + i K(du\wedge d\bar{v} - d\bar{u}\wedge dv)\right].
\ee
This is the source for the only non-vanishing component of the Riemann tensor
\be
R_{++}= {e^{-\Phi_0}\over 4}
  (G_3)_{+ij}\,(\overline{G}_3)_{+}^{\;\,ij} = 8m_0^2{e^{\Phi_0}P^2\over A_0} (16F^2+2K^2+1).
\ee
From the metric one obtains
\be
R_{++} = m_0^2\left[\,24a_2-96q_2+16f_2+8Y+2\,\right] = 8m_0^2{e^{\Phi_0}P^2\over A_0}(16F^2 + 2K^2 +1)  
\ee
after having used the relations (\ref{intcoef}). This is a check of the fact that the Einstein equation are consistently satisfied. Note that in the supersymmetric case, using (\ref{sulim1}) and (\ref{sulim2}) we recover the result in \cite{pandostras} $R_{++}= 12m_0^2(a_1/a_0)$.

The Hamiltonian and light-cone momentum look the same as the supersymmetric ones
\be
\label{KSH}
 H=i\big[\partial_t  + m_0 \,(\partial_{\phi'}+
\partial_{\psi'} - \partial_{\phi}) \big]\equiv E-m_0J,
\ee
\be
P^+= - {i\over L^2}\, m_0\,(\partial_{\phi'}+
\partial_{\psi'} - \partial_{\phi})= m_0\,\left(\,{J \over
L^2}\,\right).
\ee
There is also the additional symmetry
\be
J_A = -i(\partial_{\phi'} - \partial_{\psi'} + \partial_{\phi})  .
\ee

%%%%%%%%%%%%%%%%%%%%%%%%%%%%%%%%%%%%%%%%%%%%%%%%%%%%%%%%%%%%%%%
\subsection{String theory on the bKS pp-wave}
%%%%%%%%%%%%%%%%%%%%%%%%%%%%%%%%%%%%%%%%%%%%%%%%%%%%%%%%%%%%%%%%

We will now study the string theory on the pp-wave background of the previous Section.
The string theory is again solvable and one could repeat the same steps in \cite{pandostras}.

Let us define $m=m_0\,\alpha'\,p^+$ and
\be
m_z = \a m, \qquad  m_u = \b m, \qquad m_v=m, \qquad m_B= 4\gamma Km.
\ee 
The bosonic worldsheet fields have frequencies
\be
\label{freq1}
\omega_n^i=n, \quad i=1,2,3; \qquad\qquad \omega_n^z = \sqrt{n^2 + m_z^2};
\ee
\be
\omega_n^{u,v}=\sqrt{n^2 +{(m_u^2 +m_v^2)\over2}\pm \sqrt{{(m_u^2-m_v^2)^2\over4}+n^2 m_B^2}} .
\label{freq2}
\ee
The fermionic ones are
\be
\omega_n^{k} = \sqrt{n^2 + {m_B^2\over16K^2}(4F+1)^2}, \qquad k=1,...,4;
\label{ksfermi1}
\ee
\be
\omega_n^{\pm,l} = |\sqrt{n^2 + {m_B^2\over16K^2}(4F-1)^2} \pm {m_B\over2}|, \qquad l=1,2.\label{ksfermi2}
 \ee
Several comments are in order. Firstly, both in the broken and in the unbroken case, the sum of the squares of the bosonic frequencies equals the one of the fermionic frequencies, allowing the corresponding string model to remain finite, as observed in \cite{pandostras}.

Secondly, as in the bMN case, the bosonic sector of the broken model retains the global structure of the supersymmetric one.
The fields $v,\,\bar v$, with mass as in the supersymmetric case, represent the oscillations transverse to the
geodesic direction on the three-sphere of the deformed conifold, while the modes $u,\,\bar u,\, z$ depend on the details of the solution and
their masses depend on the supersymmetry breaking parameters.

For what concerns the world-sheet supersymmetry, again it is not linearly realized.
Nevertheless, in the original KS case the Hamiltonian after the Penrose limit
still commutes with the original four supersymmetries of the theory.
As a consequence, one finds two fermionic, massless zero-modes, as can be checked from
formula (\ref{ksfermi2}) for the supersymmetric values of the parameters.
On the contrary, in the bKS solution, i.e. for generic values of the parameters,
there are no null fermionic zero-frequencies, unless\footnote{Linking with suitable UV asymptotics of the bKS IR solutions requires $K,\,F,\, G\,$ to be positive parameters.}
\be
|1-4F|=2K.
\ee

Since there are no linearly realized worldsheet supersymmetries, the frequencies of the bosonic
and fermionic fields are different and there is in particular a non vanishing zero point energy.
The ground state energy in the bKS model reads
\be\label{zeroKS}
E_0(m)={m_0\over 2m}\sum_{n=-\infty}^{\infty}\left[3n + \omega_n^z + 2\omega_n^u + 2\omega_n^v -4\omega_n^1 -2\omega_n^{1,+}-2\omega_n^{1,-}\right].
\ee
As in the bMN case, in the $m\ra0$ limit it is easily evaluable, the main contribution coming from the zero-frequencies
\be
E_0\ra m_0\Biggl[{|\a|\over2} +|\b| +1 - 2\gamma|4F+1| - \gamma||1-4F|+2K| - \gamma||1-4F|-2K|\Biggr].
\ee 
This is a generically negative quantity. In the supersymmetric case it reads ($a_1/a_0\approx 1/4$)
\be
E_0\ra m_0\Biggl[{1\over2}\sqrt{4{a_1\over a_0}-{4\over5}} + \sqrt{4{a_1\over a_0}-{3\over5}} +1 -3\Biggl({2a_1\over a_0}\Biggr)^{1/2}\Biggr]\approx -0,26 m_0 .
\ee
Thus in the limit we find a negative zero point energy also for the supersymmetric solution, just as in the MN case.
The same considerations on the stability made for the MN case apply.

In the large $m$ limit we can approximate the sum in (\ref{zeroKS}) by an integral as it was done in the bMN case. Again it results that $E_0$ is generically negative, linearly increasing (in absolute value) with $m$. For example, in the unbroken case we find, numerically
\be
E_0\approx -{m_0m\over2}0.03434 .
\ee
Just as in the bMN case it is possible to explore here the finite temperature regime in the bKS theory. The expected implicit expression for the Hagedorn temperature has the general form (\ref{hagmn}), where the total energy is evaluated with the same prescriptions as the bMN case, in terms of the bKS masses and as usual with antiperiodic fermions. 

%%%%%%%%%%%%%%%%%%%%%%%%%%%%%%%%%%%%%%%%%%%%%%%%%%%%%%%%%%%%%%%
\section{Annulons in non supersymmetric theories}
%%%%%%%%%%%%%%%%%%%%%%%%%%%%%%%%%%%%%%%%%%%%%%%%%%%%%%%%%%%%%%%

In this Section we will discuss some topics of the field theory duals of the above strings.
We will first analyze the dual to the bMN string and then briefly review the bKS case, which is almost identical to the supersymmetric one.

%%%%%%%%%%%%%%%%%%%%%%%%%%%%%%%%%%%%%%%%%%%%%%%%%%%%%%%%%%%%%%%
\subsection{The (softly broken) MN}

Let's recall that the zero-mode spectrum of the bMN string consists of 4 massless bosons, $z,\, x_{i=1,2,3}$, 2 bosons $v_{1,2}$ of energy $m_0$, two bosons $u_{1,2}$ of energy $m_0b/2$, 4 fermions of energy $m_0(1-b/2)/2$ and 4 of energy $m_0(1+b/2)/2$.
The $x_i,\, v_j$ sector is called the universal one, the $z, \, u_j$ the non-universal one.
We would like to identify the operators dual to these modes in the field theory.
These will surely be massive hadrons built with massive KK fields, since the gauge degrees of freedom are not charged under the current $J$.
The lowest KK modes in the (b)MN solution are two massive complex chiral multiplets transforming in the bifundamental representation of $SU(2)_l\times SU(2)_r$ and in the adjoint representation of the gauge group. 
We name the four (complex) scalars of these chiral multiplets as $A_{ij}$, where the two indices $i,j=\pm$ refer to the two fundamental representations (i.e. $A_{++}$ is in the fundamental of  $SU(2)_l$ and in the anti-fundamental of $SU(2)_r$ and so on). 
The following charge assignment will help in the identification:
\vskip 0.5truecm
\begin{center}
{\small
\begin{tabular}{|c|c|c|c|c|}
\hline
& $A_{++}$ & $A_{+-}$ & $A_{-+}$ & $A_{--}$ \\
\hline
$E$ &  $m_0$  & $m_0$   & $m_0$   & $m_0$  \\
\hline
$U(1)_l$ & 1/2  & 1/2 & -1/2 & -1/2 \\
\hline
$U(1)_{J_1}$ & 0  & 0 & 0 & 0 \\
\hline
$U(1)_\psi$ & -1/2  & 1/2 & -1/2 & 1/2 \\ \hline
$J$  & 0  & 1 & -1 & 0  \\
\hline
\end{tabular}}
\vskip 0.3truecm
\end{center}
Recall that $J=U(1)_l -{b \over 2} U(1)_{J_1} + U(1)_\psi$.
These charges are justified as follows. 
The masses (``$E$'' in the table above) are all of order $m_0$ in the (b)MN theory. 
The precise coefficients are not known in field theory, since these are the masses as computed in the dual string theory, i.e. they are the masses at strong coupling in field theory.
We will see that these values reproduce the string theory expectations, so that they can be viewed as the string prediction for the masses\footnote{Recall that the value of $m_0$ is arbitrary in the supergravity solution, so what we mean here is that the string theory predicts the right ratios of the hadron masses.}.
The $U(1)_l$ charges are the ones described above, and the $U(1)_\psi$ charges are the $U(1)_r$ ones, as $\psi$ is the R-symmetry in the MN solution, precisely given by the (twisted) $U(1)_r\subset SU(2)_r$. 
The $U(1)_{J_1}$ are all zero, these scalars being all uncharged with respect to the (2-sphere) spin connection\footnote{This is the reason why in the \none solution in the wrapped five-brane constructions there are no massless scalars, as their $U(1)_{twist}=U(1)_r$ is non-zero, while in the \ntwo case there are two, since for two of them the $U(1)_{twist}=U(1)_r+U(1)_l$ is zero \cite{noi2}.}.

Let's make a tentative identification of the string zero modes with operators in the universal sector.
Since the geodesic used to perform the Penrose limit was on the equator of the three sphere, it is natural to identify that direction with the vacuum of the string theory, so that we can guess that the latter is dual to $(A_{+-})^{J}$ acting on the field theory vacuum ($A_{-+}^{\dagger}$ is degenerate with $A_{+-}$ and there can be such fields too in the vacuum).
Its energy is $Jm_0$ and since each constituent has $J=1$, one ends up with\footnote{We will discuss this value at the end of this Section.} $H=E-m_0J=0$.
The oscillations transverse to the geodesic are naturally identified with $v, \, \bar v$ ($v=v_1 + iv_2$). 
The dual operators are then (the insertion in the string of $A_{+-}$ of) $A_{++}$ and $A_{--}$, which are conjugate to each other and whose Hamiltonian is $H=E-m_0J=m_0-m_0*0=m_0$, the value of the corresponding string oscillators.
Also these operators are degenerate with their (interchanged) conjugates.
The remaining $A$ operators, namely $A_{+-}^{\dagger}$ and $A_{-+}$, could be obtained from the vacuum with two actions of the $J$ current.
But this transformation would be subleading in $1/J$ with respect to $J$ giving the two modes $A_{++}$ and $A_{--}$, so $A_{+-}^{\dagger}$ and $A_{-+}$ are expected to be unstable and decay to stable modes \cite{pandostras}.
Finally, the $x_i$ modes are interpreted as in the supersymmetric case \cite{pandostras}.
The hadrons we are talking about are really heavy, string-like objects, hence the name {\it Annulons}, moving in an essential non-relativistic way in the three special directions.
As such, the motion and excitations in these directions will be just like three ordinary, flat-background string modes, the $x_i$ indeed. 

Let's now look at the charges of the fermionic superpartners of the $A_{ij}$:
\vskip 0.5truecm
\begin{center}
{\small
\begin{tabular}{|c|c|c|c|c|}
\hline
&  $\psi_{+}^{+}$  & $\psi_{+}^{-}$ &  $\psi_{-}^{+}$ &  $\psi_{-}^{-}$ \\
\hline
$E$ &  $m_0$  & $m_0$   & $m_0$   & $m_0$  \\
\hline
$U(1)_l$ & 1/2  & 1/2 & -1/2 & -1/2  \\
\hline
$U(1)_{J_1}$ & 1/2  & -1/2 & 1/2 & -1/2  \\
\hline
$U(1)_\psi$ & 0 & 0 & 0 & 0  \\
\hline
$J_{susy}$  & 1/3  & 2/3 & -2/3 & -1/3 \\
\hline
$J$  & (1-b/2)/2  & (1+b/2)/2  & -(1+b/2)/2 & -(1-b/2)/2  \\
\hline
\end{tabular}}
\vskip 0.3truecm
\end{center}
Here the lower and higher indices denote the $U(1)_l$ and $U(1)_{J_1}$ charges.
In the (b)MN theory, the four field theory fermions coming from the reduction from six to four dimensions are in the ({\bf 2,0}) and ({\bf 0,2}) of $SU(2)_l\times SU(2)_r$.
Upon twisting, one of the two charged under $SU(2)_r$ stays massless, giving the gaugino of the \none theory, the other one becomes very massive.
The other two fermions are massive and give precisely the fields in the table above, hence the $U(1)_l$ and $U(1)_{\psi}$ charges.
The $U(1)_{J_1}$ charges come from the fact that these fermions are charged under the spin connection.
Finally, these fermions have all masses equal to the ones in the supersymmetric case.
While this could not have been the case, since we have broken supersymmetry, it turns out that the identification with the string states is straightforward.
Moreover, it is likely that the difference in mass between bosons and fermions is of order $1/J$, thus it is not seen in the Penrose limit.

The spinor $\psi_{+}^{+}$ has $H=m_0-m_0*(1-b/2)/2=m_0*(1+b/2)/2$. 
In the same way one reads that the same result holds for $\bar{\psi}_{-}^{-}$, while $\psi_{+}^{-}$ and  $\bar{\psi}_{-}^{+}$ have $H=m_0*(1-b/2)/2$.
We see that we have in a row the four string fermionic modes with energy $m_0*(1+b/2)/2$ and four with energy $m_0*(1-b/2)/2$.

For what concerns the non universal sector, i.e. the fields $z,\, u_1,\, u_2$, the identification is not so straightforward.
There are many KK states in the theory which could account for these modes, but since we don't have a completely definite proposal, we won't go further in this sector.

Finally, let's comment on the value of the vacuum state Hamiltonian as found above, namely zero.
It is {\it not} the value predicted by the string theory.
In the latter case, as we have seen, the vacuum energy has a negative value, with a linear dependence on the parameter $m=m_0\alpha'p^+$ in the large $m$ limit.
This is the limit in which the dual field theory should be effectively weakly coupled.
Then the string theory predicts a value for $H$ which is zero perturbatively in the field theory effective coupling constant, {\it but which has corrections in inverse powers of this coupling}.
The situation is very similar to the Type 0 case \cite{noi0} in spirit but not in practice.
There, the vacuum energy in the twisted sector was negative but exponentially vanishing in the large $m$ regime, allowing for the use of the perturbative expansion for the field theory calculations of the anomalous dimensions.
Here, instead, the ``non perturbative'' corrections are power-like and large.
As for the Type 0 case, we don't have a field theory explanation for this behavior, which would involve the estimation of some kind of non perturbative contribution in the effective coupling constant for the annulons, i.e. in the large energy, large $J$ regime.
We leave this interesting subject for future investigation.

%%%%%%%%%%%%%%%%%%%%%%%%%%%%%%%%%%%%%%%%%%%%%%%%%%%%%%%%%
\subsection{The (softly broken) KS}

Let's consider now the (b)KS case and briefly recall what can be said about the correspondence between the string modes and field theory hadrons \cite{pandostras}.
The KS model is dual, in the regime we are considering, to a \none SYM theory with gauge group $SU(M)$, reached after a cascade of Seiberg dualities from a theory with gauge group $SU(N+M)\times SU(N)$, bifundamental chiral multiplets and a superpotential.
The $SU(M)$ gauge theory is coupled to many fields coming as products of the duality cascade.
Among the many vacua of the theory there are the ones in which mesonic fields $N_{ij}$, $i,j=1,2$, acquire mass.
These are the right fields to look at since they carry the right charges under the currents $J$ and $J_A$ introduced in Section 4, while, for example, the gauge fields are uncharged under them.
Under ($J$,$J_A$), $N_{11}$ has charges ($1,0$), $N_{22}$ has charges ($-1,0$), while $N_{12}$ and $N_{21}$ have charges ($0,1$) and ($0,-1$) respectively.
Then, the most natural candidate for the string ground state is the operator\footnote{Also in this case there are degeneracies with the complex conjugate fields.} $(N_{11})^J$ acting on the field theory vacuum, since it is uncharged under $J_A$, which represents the directions orthogonal to the geodesic used in the Penrose limit, and, if we suppose that the $N_{ij}$-meson mass is exactly $m_0$, it has zero\footnote{The same considerations made for the bMN theory on this value of $H$ apply in this case.} Hamiltonian $H=E-m_0*J=0$.
The value $m_0$ for the mass of the mesons can be seen as the string theory prediction in the strong coupling regime of the field theory.
The two modes transverse to the geodesic, $v,\, \bar v$, are naturally identified with the mesons charged under $J_A$, namely $N_{12}$ and $N_{21}$, which are in fact conjugate to each other.
They have $H=m_0-m_0*0=m_0$, the value predicted by the string.
The $N_{22}$ field, instead, is believed to be absent in the large $J$ limit, just like $A_{-+}$ in the bMN theory.
Finally, the three $x_i$ directions give again the non relativistic motion of the annulons in the three special directions.
The universal sector is then very similar to the (b)MN one also from the field theory point of view. 
In particular, it is not affected by the supersymmetry breaking, so that we can repeat the above identifications verbatim in the bKS theory.

The fermionic and the non universal sectors are quite non-trivial to identify even in the supersymmetric case, not to talk about the five parameter dependent bKS theory.
But in the case of the fermionic superpartner of $N_{11}$, called $\psi_{11}$ in \cite{pandostras}, we have a prediction for the mass from the string theory.
In fact, in the supersymmetric case it is associated with the two fermionic zero-modes of the string theory, since its $J$, $J_A$ quantum numbers and its mass are the same of $N_{11}$, giving $H=0$ again.
In the softly broken theory we can recognize these modes as the ones in (\ref{ksfermi2}), with Hamiltonian $H=m_0\gamma||4F-1|-2K|$.
Since the $J$, $J_A$ quantum numbers of $\psi_{11}$ are not changed by the supersymmetry breaking, we conclude that its mass is modified, as expected, and it is equal to $m_0[\gamma||4F-1|-2K|+1]$.
In particular, the ``flat directions'' with $K=|4F-1|/2$ give the same mass as in the supersymmetric case, restoring the bosonic/fermionic degeneracy for this modes.
As in the bMN case, this degeneracy in the broken theory is likely to be an accidental feature of the large $J$ limit and it is expected to be absent if the subleading contributions in $1/J$ are considered. \\

\begin{center}
{\large  {\bf Acknowledgments}}
\end{center}
We thank R. Casero, L. Pando-Zayas and A. Zaffaroni for useful discussions. F. B. is partially supported by INFN.

%%%%%%%%%%%%%%%%%%%%%%%%%%%%%%%%%%%%%%%%%%%%%%%%%%%%%%%%%%%%%%%%%%%%%%%%%%%%%%%%%%%%%%%%%%%%%%%%%%%%%%%

\end{document}